# Compensation-dependent in-plane magnetization reversal processes in $Ga_{1-x}Mn_xP_{1-y}S_y$

P.R. Stone[1,2*], C. Bihler[3], M. Kraus[3], M.A. Scarpulla[1,2], J.W. Beeman[2], K.M. Yu[2], M.S. Brandt[3] and O.D. Dubon[1,2**]

[1]*Department of Materials Science & Engineering, University of California, Berkeley, CA 94720*
[2]*Lawrence Berkeley National Laboratory, Berkeley, CA 94720*
[3]*Walter Schottky Institut, Technische Universität München, Am Coulombwall 3, 85748 Garching, Germany*

ABSTRACT

We report the effect of dilute alloying of the anion sublattice with S on the in-plane uniaxial magnetic anisotropy and magnetization reversal process in $Ga_{1-x}Mn_xP$ as measured by both ferromagnetic resonance (FMR) and superconducting quantum interference device (SQUID) magnetometry. At $T$=5K, raising the S concentration increases the uniaxial magnetic anisotropy between in-plane <011> directions while decreasing the magnitude of the (negative) cubic anisotropy field. Simulation of the SQUID magnetometry indicates that the energy required for the nucleation and growth of domain walls decreases with increasing $y$. These combined effects have a marked influence on the shape of the field-dependent magnetization curves; while the $[0\bar{1}1]$ direction remains the easy axis in the plane of the film, the field dependence of the magnetization develops double hysteresis loops in the [011] direction as the S concentration increases similar to those observed for perpendicular magnetization reversal in lightly doped $Ga_{1-x}Mn_xAs$. The incidence of double hysteresis loops is explained with a simple model whereby magnetization reversal occurs by a combination of coherent spin rotation and noncoherent spin switching, which is consistent with both FMR and magnetometry experiments. The evolution of magnetic properties with S concentration is attributed to compensation of Mn acceptors by S donors, which results in a lowering of the concentration of holes that mediate ferromagnetism.

___________________________________________________________________________

[*]Electronic mail: prstone@berkeley.edu
[**]Electronic Mail: oddubon@berkeley.edu

## I. INTRODUCTION

Because of their potential as both injectors and filters for spin-polarized carriers, ferromagnetic semiconductors have been proposed for use in spin-based electronics, or spintronics [1, 2]. One particularly well-studied class of ferromagnetic semiconductors is synthesized by replacing a few percent of the cation sublattice in a III-V semiconductor host with Mn. Since the dilute Mn moments are spatially separated, ferromagnetic exchange in III$_{1-x}$Mn$_x$V ferromagnetic semiconductors is indirect and mediated by holes provided by the substitutional Mn acceptors. The resulting strong correlation between the spin-polarized holes and Mn moments causes the magnetic anisotropy to be strongly dependent on the nature of the mediating carriers. Previous studies, carried out predominantly on the canonical Ga$_{1-x}$Mn$_x$As system, have demonstrated control of the orientation of the magnetic easy axis by modulation of the carrier concentration [3], epitaxial strain [4], and temperature [5].

To date, theoretical understanding of magnetic anisotropy in III$_{1-x}$Mn$_x$V systems has focused mainly on employing the Zener model, which assumes that ferromagnetism is mediated by itinerant holes of mostly valence-band character [3, 6-9]. It is therefore enlightening to examine the magnetic anisotropy in a material in which exchange is unambiguously mediated by more localized holes of impurity band character, e.g. Ga$_{1-x}$Mn$_x$P [10]. We have previously reported on several aspects of the magnetic anisotropy in Ga$_{1-x}$Mn$_x$P. As is the case with Ga$_{1-x}$Mn$_x$As we find that growing Ga$_{1-x}$Mn$_x$P films in tensile strain results in a magnetic easy axis that is perpendicular to the thin-film plane while films grown in compressive strain exhibit in-plane easy axes [11]. The combination of ferromagnetic resonance (FMR) spectroscopy and SQUID magnetometry indicates that Ga$_{0.958}$Mn$_{0.042}$P films grown on GaP exhibit a cubic anisotropy contribution with a sign opposite to the one commonly observed for Ga$_{1-x}$Mn$_x$As [12].

There also is a uniaxial component to the in-plane anisotropy wherein the magnetically easy $[01\bar{1}]$ direction is inequivalent to [011] [12]. This in-plane component to the magnetic anisotropy seems general to $III_{1-x}Mn_xV$ systems as it is also observed in $In_{1-x}Mn_xAs$ [13] and $Ga_{1-x}Mn_xAs$ [3, 5, 14]. In $Ga_{1-x}Mn_xAs$ it has been demonstrated that the magnitude and sign of this in-plane uniaxial anisotropy field is intimately related to the carrier concentration [3].

In this work we report the effect of compensation on the in-plane magnetic anisotropy and magnetization reversal process in $Ga_{1-x}Mn_xP_{1-y}S_y$. The in-plane uniaxial anisotropy is found to increase with increasing $y$, i.e. with *decreasing* hole concentration, $p$. By applying a simple model [15] that takes into account both coherent Stoner-Wohlfarth-like rotation of the magnetic moments and noncoherent spin switching, we extract formation energies for both 90 and 180 degree domain walls and find that these energies decrease with increasing $y$. The combination of these two effects results in the development of double hysteresis loops when the magnetic field is applied parallel to [011] as $y$ is increased.

## II. EXPERIMENTAL

All samples were synthesized using the combination of ion implantation and pulsed-laser melting (II-PLM) [16, 17]. $Ga_{1-x}Mn_xP$ was synthesized by implantation of 50 keV $Mn^+$ into (100)-oriented GaP to a dose of $1.5 \times 10^{16}$ cm$^{-2}$ followed by irradiation with a single pulse from a KrF ($\lambda$=248 nm) excimer laser at a fluence of 0.44 J/cm$^2$. Quaternary alloys were synthesized by co-implanting the Mn-implanted GaP with 60 keV $S^+$ to doses ranging from 2.5 to 7.3 $\times 10^{15}$ cm$^{-2}$ prior to PLM. Prolonged (24 hours) etching in concentrated HCl was used to remove a very thin highly defective surface layer as well as any surface oxide phases [16, 18].

Selected compositional parameters for the samples used in this study are presented in Table 1. The substitutional Mn concentration was determined by the combination of secondary ion mass spectrometry (SIMS) and $^4$He$^+$ ion beam analysis[12, 16]. We define $x$ as the peak Mn$_{Ga}$ concentration [19]. For samples used in this study we observe a substitutional fraction ($f_{sub}$) of Mn atoms substituting for Ga of 80-88%, which is comparable to that observed in Ga$_{1-x}$Mn$_x$As thin films grown by low temperature molecular beam epitaxy (LT-MBE) [20]. More importantly the II-PLM process results in no interstitial Mn (Mn$_I$); the remainder of the Mn atoms is incommensurate with the lattice. Therefore, II-PLM films have no unintentional compensation due to Mn-related defects in our films, allowing for more reliable control of the carrier concentration in this study. The *total* S concentration as a function of depth was determined by SIMS. The peak substitutional sulphur concentration, $y$, was estimated by multiplying the peak in the S concentration by the typical substitutional fractions of dopants in GaP and GaAs after PLM, which ranges between 75% and 90% based on our previous pulsed-laser melting studies of dopant incorporation into GaP and GaAs [10, 17, 21]. We note that $y$ refers to the concentration of *substitutional* sulphur. The fraction of S atoms that are *electrically active* was estimated from Hall effect measurements on GaP$_{0.979}$S$_{0.021}$ synthesized under identical conditions to the Ga$_{0.959}$Mn$_{0.041}$P$_{0.979}$S$_{0.021}$ film. Without Mn the carrier concentration can be measured as the anomalous Hall contribution to the Hall resistivity is absent. Comparison of the Hall effect, SIMS, and ion channeling data indicate that ~36% of substituionally incorporated S atoms are electrically active. Ion channeling measurements on S-doped Ga$_{1-x}$Mn$_x$As show that there is no appreciable incorporation of S into interstitial sites. Given the similarities in processing between GaAs and GaP this allows us to preclude the formation of interstitial S defects in the samples presented in this work.

DC magnetization measurements were performed using a SQUID magnetometer. FMR measurements were performed at $\omega/2\pi \approx 9.26$ GHz in an electron paramagnetic resonance (EPR) spectrometer using magnetic field modulation with the sample temperature controlled using a liquid-He flow cryostat. Differentiation of the in-plane <011> directions was accomplished by etching samples in $H_3PO_4$ at 180° C for 5 minutes, which produces asymmetric etch pits oriented in the $[0\bar{1}1]$ direction [22, 23].

## III. RESULTS AND DISCUSSION

### A. Ferromagnetic Resonance Spectroscopy

The field dependence of the FMR intensity measured at $T$=5K and with the field parallel to the in-plane $[0\bar{1}1]$ crystallographic direction for various S concentrations is shown in Figure 1. In the absence of sulphur, there is a single, orientation-dependent resonance associated with the collective mode of the ferromagnetically-coupled Mn moments. As $y$ increases, a second feature emerges at $\mu_o H \sim 330$ mT that shows only a slight angular dependence. For the external magnetic field oriented perpendicular to the film plane the resonance is located at $\mu_o H_{res} = 330.3$ mT, which corresponds to a $g$-factor of 2.004. This value for the $g$-factor is identical to that observed for paramagnetic $Mn^{2+}$ in GaP [24]. The peak-to-peak linewidth $\mu_o H_{pp}$=11.8mT of the resonance can be explained by a broadening of the underlying hyperfine structure in analogy to Ref. [25]. For the external magnetic field aligned within the film plane the resonance position is shifted by ~1.5mT to $\mu_o H_{res} \sim 328.8$ mT ($g$=2.013). A similar effective shift of the resonance field in highly Mn doped GaAs has already been reported for the ionized Mn acceptor in $Ga_{1-}$ $_x$Mn$_x$As [25]. The authors of Ref. [25] attributed the slight shift in the resonance position to demagnetizing field effects of Mn$^{2+}$ ions at low temperatures. In principle the paramagnetic resonance could be caused by electrons at neutral S donors. However, we exclude this possibility due to the differing *g*-factor *g*=1.998 and the smaller line width $\mu_o H_{pp}$~6mT [26, 27]. We therefore attribute the resonance at ~330mT to paramagnetic Mn$^{2+}$.

As *y* increases the magnitude of the paramagnetic Mn$^{2+}$ resonance increases with respect to that of the collective mode. Such behavior is consistent with a model in which compensation of ferromagnetism-mediating holes by electrically active S donors decouples an increasing number of Mn moments from ferromagnetic exchange. For larger *y* the inhomogeneous distribution of Mn and S throughout the film thickness gives rise to regions of the film where the S concentration is greater than or equal to the Mn concentration, as is shown in Figure 2. In such regions there is an insufficient concentration of holes to transmit spin information among nearest neighbor Mn atoms due to strong compensation by sulphur donors which makes such regions paramagnetic. It should be noted that the region of the film with the peak Mn concentration has an adequate hole concentration to support extended ferromagnetic exchange. Indeed, Ga$_{0.959}$Mn$_{0.041}$P$_{0.973}$S$_{0.027}$ has a Curie temperature ($T_C$) of 21 K [28, 29]. Hence at *T*=5K the FMR measurement detects both the resonance of the ferromagnetically-coupled Mn atoms as well as that of the paramagnetic Mn moments. As *y* increases, the total amount of atoms in the paramagnetic "tail" of the Mn distribution increases, as the depth over which the sulphur concentration is sufficient to completely disrupt ferromagnetic exchanges is larger. Consequently, the relative intensity of the paramagnetic resonance to the collective ferromagnetic resonance increases with increasing *y*.

The decrease in the fraction of ferromagnetically coupled Mn atoms with increasing $y$ is supported by SQUID magnetometry measurements. The magnetic moment at $\mu_0 H=5T$ decreases monotonically from $3.5\pm0.2$ $\mu_B/Mn_{Ga}$ for $y=0$ to $1.9\pm0.2$ $\mu_B/Mn_{Ga}$ for $y=0.027$ since at this magnitude of the applied field the paramagnetic Mn spins are not yet completely aligned along the field direction. Further support for sulphur-induced compensation comes from the decrease of both $T_C$ and the XMCD asymmetry decrease with increasing $y$ [28, 29], similarity to results obtained in Te co-doped $Ga_{1-x}Mn_xP$ [10], and aforementioned Hall measurements, which demonstrated an electrical activation of ~36% of substituitonal S donors in GaP:S synthesized by II-PLM.

As we are primarily concerned with examining the in-plane magnetization reversal processes in the ferromagnetic region of the film the remainder of the data analysis will focus on the resonance of the collective mode. Panels (a), (b), and (c) of Figure 3 show the angular dependence of the resonance field for rotation about the [100] axis, i.e. for various magnetic field orientations in the plane of the sample. For all $y$ the in-plane rotations are characterized by local minima occurring when the applied field is parallel to <011> directions and maxima when the field is parallel to <001> directions. The rotations do not exhibit four-fold symmetry. The resonance fields at $H\|[0\bar{1}1]$ are smaller in magnitude than those at $H\|[011]$ indicating that an in-plane uniaxial component to the magnetic anisotropy is present in $Ga_{1-x}Mn_xAs_{1-y}S_y$ for all $y$.

We examine the in-plane magnetic anisotropy in more detail by simulating the in-plane angular dependence of the FMR. We write the free energy as a function of the orientations of the applied magnetic field and sample magnetization vector:

$$F = -MH(\sin\Theta\sin\Phi\sin\theta\sin\phi + \cos\Theta\cos\theta + \sin\Theta\cos\Phi\sin\theta\cos\phi)$$
$$+ K_{eff}^{100}\sin^2\Theta\sin^2\Phi - \frac{1}{2}K_{c1}^{\perp}\sin^4\Theta\sin^4\Phi \qquad (1)$$
$$- \frac{1}{2}K_{c1}^{\parallel}(\cos^4\Theta + \sin^4\Theta\cos^4\Phi) + \frac{1}{2}K_u^{011}(\cos\Theta + \sin\Theta\cos\Phi)^2$$

In Equation (1) $\theta$, $\phi$ and $\Theta$, $\Phi$ define the orientation of the magnetic field and magnetization vectors, respectively, according to the coordinate system shown in Figure 3 (d). The first term in Equation (1) is the Zeeman energy. The second term represents the total out-of-plane uniaxial anisotropy field, $K_{eff}^{100}$, which is the sum of magnetocrystalline and shape effects. The third and fourth terms represent the out-of-plane and in-plane cubic anisotropy contributions, which arise due to the tetragonal distortion of the epitaxial film [30]. The final term, $K_u^{011}$, is included to account for the aforementioned uniaxial anisotropy that exists between in-plane [011] and $[01\bar{1}]$ directions.

Following the approach of Smit *et al.* [31, 32] we obtain the equation of motion

$$\left(\frac{\omega}{\gamma}\right)^2 = \frac{1}{M^2\sin^2\Theta}\left[\left(\frac{\partial^2}{d\Phi^2}F\right)\left(\frac{\partial^2}{\partial\Theta^2}F\right) - \left(\frac{\partial}{\partial\Phi}\frac{\partial}{\partial\Theta}F\right)^2\right]_{\Phi_0,\Theta_0} \qquad (2)$$

where $\gamma = g\mu_B/\hbar$ is the gyromagnetic ratio. Equation (2) is evaluated at the equilibrium orientation of the magnetization, which is found by the minimization conditions,

$$\frac{\partial}{\partial\Phi}F\bigg|_{\Phi=\Phi_0} = \frac{\partial}{\partial\Theta}F\bigg|_{\Theta=\Theta_0} = 0. \qquad (3)$$

The simultaneous solution of Equations (2) and (3) yields the FMR resonance condition at a specific magnetic field orientation. The inputs to Equations (2) and (3) are the orientation of the applied field, the microwave frequency, the *g*-factor, and the various anisotropy fields. The solution of this system of equations results in the equilibrium orientation of magnetization and the resonance field. In the fitting process the set of parameters is successively adapted to reproduce the measured angular dependence of the resonance field.

The solid lines in Figure 3 (a), (b), and (c) are simulations of the angular dependence of the FMR, which were found to agree with the out-of-plane FMR rotations as well (data not shown) for the range of anisotropy fields listed in Table 2. The out-of-plane cubic and uniaxial anisotropy fields dominate the magnetic anisotropy in all samples. The combined effect of these two terms leads to a strong preference for the magnetization vector to lie in the plane of the sample as opposed to an arbitrary out-of-plane orientation. In general, uniaxial ($2K_{eff}^{100}/M$) and cubic ($2K_c^{\perp}/M$) anisotropy contributions perpendicular to the film plane are smaller in magnitude in the compensated materials, though the trend is not monotonic. The origin of this quantitative behavior is not completely understood at this time and will not be further discussed. We point out only that these values of the out-of-plane anisotropy fields influence the absolute magnitude of the in-plane resonance fields shown in Figures 1 and 3. Hence the resonance field for $H \| \left[0\bar{1}1\right]$ occurs at a lower field for $y=0$ than for $y=0.010$, 0.021 or 0.027 due to the enhanced out-of-plane anisotropy for the former material. On the other hand, the relative difference in resonance fields for different in-plane magnetic field orientations within a given sample (quantified by $K_{c1}^{\|}$ and $K_u^{011}$) will not be significantly affected by this effect, which allows us to compare the relative evolution of the resonance fields for in-plane rotations of the external magnetic field between different materials. For all $y$, $K_{c1}^{\|}$ is negative while $K_u^{011}$ is positive. The former results in the <011>-type directions being magnetically preferred over <001>-type directions while the latter determines that the in-plane easy axis is oriented parallel to $\left[0\bar{1}1\right]$ as opposed to [011]. Increasing $y$ results in significant enhancement of $K_u^{011}$ indicating that the inequivalence between the $[0\bar{1}1]$ and [011] in-plane directions grows as $y$ increases. This is consistent with the behavior observed in $Ga_{1-x}Mn_xAs$ in which a rotation of the easy axis from $[0\bar{1}1]$ to [011] occurs upon increasing $p$ [3]. Here we observe a similar trend. $Ga_{0.958}Mn_{0.042}P$ starts with an $[0\bar{1}1]$ easy axis which is increasingly preferred over [011] as $y$ increases, and $p$ *decreases*.

**B. SQUID Magnetometry Measurements**

The effect of the changes in the in-plane magnetic anisotropy fields on the process by which in-plane magnetization reversal occurs will now be explored. Figure 4 shows the field dependence of the magnetization, $M(H)$, at $T=5K$ for the applied external field oriented parallel to either the [011] or $[0\bar{1}1]$ direction as measured by SQUID magnetometry. For all $y$ the $M(H)$ curves measured with $H\|[0\bar{1}1]$ behave in a qualitatively similar manner. All exhibit relatively low coercivities (less than 4 mT) and exhibit easy-axis square-like hysteresis loops in agreement with the in-plane FMR rotations. As $y$ increases, the remnant magnetization gets progressively lower. This is not a result of a magnetic hardening of the $[0\bar{1}1]$ direction but is instead the result of the previously discussed decoupling of $Mn_{Ga}$ moments from the global inter-Mn exchange interaction as $y$ increases.

The behavior of the *H*∥[011] *M*(*H*) curves with increasing *y* is markedly different as the increase in the magnitude of $K_u^{011}$ has a significant effect on magnetization reversal process. For *y*=0 the *M*(*H*) curve has a kink near zero applied field, which has previously been attributed to a multistep process in which both coherent spin rotation and noncoherent spin switching mechanisms are operative and will be discussed in more detail below [15]. As *y* increases the kinked *M*(*H*) curve evolves into a "wasp-waisted" or "double" hysteresis loop. Double hysteresis loops have previously been observed in Ga$_{1-x}$Mn$_x$As for the case of *perpendicular* magnetization reversal at low Mn concentrations [33]. The existence of double hysteresis loops in Ga$_{1-x}$Mn$_x$As was attributed by Titova *et al.* to the complex nature of the free energy surface arising from uniaxial and cubic anisotropy terms that were of similar magnitude [33]. For Ga$_{1-x}$Mn$_x$P$_{1-y}$S$_y$ we observe 3.5≤| $K_{c1}^{\parallel}$/ $K_u^{011}$ |≤2 for the wasp-wasited loops. In this regard, our results are consistent with this explanation.

**C. Modeling of Hysteresis Loops**

To further elucidate the factors governing the appearance of the double hysteresis loops, the in-plane *M*(*H*) curves have been simulated. Here, we consider the case where the magnetization and the applied field are confined to the plane of the thin film (ϕ=Φ=0°, or 180°). Under these conditions Equation (1) takes on the more simple form,

$$F = -MH(\cos\Theta\cos\theta + \sin\Theta\sin\theta) - \frac{1}{2}K_{c1}^{\parallel}(\cos^4\Theta + \sin^4\Theta) + \frac{1}{2}K_u^{011}(\cos\Theta + \sin\Theta)^2 . \quad (4)$$

The assumption that Φ=0/180° is reasonable given the compressive strain state of Ga$_{1-x}$Mn$_x$P$_{1-y}$S$_y$/GaP thin films, which induces the easy axis to lie in the film plane [11]. For in-plane rotations (ϕ=0/180°) the magnetic moment will be constrained to the plane of the film as both the Zeeman and anisotropy energies favor an in plane orientation of the magnetic moment for this geometry, which justifies the use of Equation (4). The calculations are performed by inputting the values for the two in-plane anisotropy fields determined from FMR, which allows for the free energy to be calculated as a function of the in-plane orientation of the magnetization $F(\Theta)$ for a given magnetic field. A free parameter, $\Delta E^{011}$ ($\Delta E^{0\bar{1}1}$) is used to account for the energy required to nucleate and grow domains that gives rise to hysteresis during the magnetization reversal process when the field is parallel to the [011] ($[0\bar{1}1]$) direction. The values of $\Delta E^{011}$ and $\Delta E^{0\bar{1}1}$ are chosen such that the measured fields at which noncoherent spin switches occur are reproduced by the simulation.

The simulated $M(H)$ curves for $y=0.010$ and $H\|[011]$ as well as the $F(\Theta)$ calculated at selected magnetic fields are shown in Figures 5(a) and (b) following the approach of Refs. [12, 15] using the anisotropy fields listed in Table 2. At high magnetic fields the Zeeman term is dominant, which results in a global minimum in the free energy landscape parallel to the field direction [011] ($\Theta=45°$). As the magnitude of the applied field is decreased towards zero, the anisotropy energy overcomes the Zeeman energy and a new global minimum emerges close to the $[0\bar{1}1]$ easy axis ($\Theta\approx-45°$). The magnetization will remain oriented parallel to [011] until the energy gained by switching from the local minimum at $\Theta=45°$ to the global minimum at $\Theta\approx-45°$ is equal to $\Delta E^{011}$. Reasonable agreement with experiment is achieved by assuming $\Delta E^{011} = 1.9\times10^{-4}$ meV/Mn, which causes the first noncoherent spin flip to occur at a field, $\mu_0 H_1=2.2$ mT. As the field is decreased through zero, magnetization reversal proceeds by coherent spin rotation due to the gradual angular shift in the minimum in $F(\Theta)$ by the Zeeman contribution to the free energy landscape. Coherent spin rotation continues until a new global energy minimum that is 1.9x10$^{-4}$ meV/Mn lower in energy emerges parallel to $[0\bar{1}\bar{1}]$. For the $y$=0.010 simulation the second noncoherent spin flip occurs at -$\mu_0H_2$=-6.0 mT. Similar arguments apply upon sweeping from negative to positive fields with noncoherent switches occurring at -$\mu_0H_1$=-2.2 and $\mu_0H_2$=6.0 mT, which completes the double hysteresis loop. Figures 6 (a) and (b) show the simulated hysteresis loop and selected $F(\Theta)$ contours for $y$=0.027. Both the decreased width of each half of the double hysteresis loop as well as the higher field at which the center of each half of the double hysteresis loop occurs are captured by our simple model using a reduced value for $\Delta E^{011}$ of 6.2x10$^{-5}$ meV/Mn and the in-plane anisotropy fields determined from FMR.

As a guide to further discussion of the SQUID simulations, Figure 7 indicates the effects of the key model parameters on the shape of the double hysteresis loops. The "width" of each half of the double hysteresis loop ($\mu_0H_2$-$\mu_0H_1$) is predominantly governed by the value of $\Delta E^{011}$. It should be noted that while the simulated hysteresis loops are in reasonable qualitative agreement with experiment, they predict much sharper noncoherent spin switching processes than are observed experimentally, which leads to a range of values of $\Delta E^{011}$ which reasonably describe the experimental results. A more rigorous quantitative analysis suggests the ranges for $\Delta E^{011}$ as a function of $y$ that are listed in Table 3. Although the spread in values is rather large the general trend still suggests that $\Delta E^{011}$ decreases with increasing $y$.

One reason for the disagreement is that our simulations account only for hysteretic effects caused by noncoherent spin switching and neglect other processes, particularly the pinning and depinning of domain walls by defects during magnetization reversal. Furthermore, our simple model assumes a single, uniform magnetic phase in which the anisotropy is adequately described by single-valued parameters. The random distribution of Mn can lead to sample inhomogeneities which give rise to local fluctuations in the magnetization and hole concentration. In fact, recent results by Kim *et al*. have demonstrated that such fluctuations give rise to a broad distribution of domain pinning fields ($\Delta E/M$) in annealed, LT-MBE grown $Ga_{1-x}Mn_xAs$ thin films which is well described by a broad Gaussian distribution [34]. We follow a similar approach here, as is shown in Figure 8 for *y*=0.010, where a Gaussian distribution of $\Delta E^{011}$ with a mean of $1.2\times10^{-4}$ meV/Mn and standard deviation of $4.9\times10^{-5}$ meV/Mn is used. Clearly, the calculated *M(H)* loops agree much better with experiment when a distribution of $\Delta E$ is used suggesting such local fluctuations in the magnetization (and as a result the magnetic anisotropy) play a signficiant role in determining the distribution of domain pinning energies in $Ga_{1-x}Mn_xP$ as well. The similar distributions of domain wall energies between the annealed LT-MBE grown $Ga_{1-x}Mn_xAs$ and II-PLM formed $Ga_{1-x}Mn_xP_{1-y}S_y$ suggests that lateral inhomogeneity is inherent to materials grown by both processes and that the vertical inhomogeneity inherent to II-PLM processing does not produce additional spread in the domain wall energies.

Assuming Gaussian distributions the switching energies for both [011]- and $[0\bar{1}1]$-oriented magnetization reversal processes have been calculated as a function of *y* and are shown in Figure 9. For all *y* $\Delta E^{011} < \Delta E^{0\bar{1}1}$, which is reasonable considering that the former case requires nucleation of 90 degree domain walls while the latter involves 180 degree domain wall formation [12]. Generally speaking, there is a decrease in $\Delta E$ as *y* increases for both [011] and $[0\bar{1}1]$ magnetization reversal indicating that noncoherent spin switching become easier (that is requires less energy) as the carrier concentration is decreased. This trend is in agreement with the rough estimates made using the single-valued $\Delta E$ model.

The slope of the near-zero field portion of $M(H)$ is in principle determined by $K_c^\parallel/M$ (Figure 7). As magnetization reversal occurs in this regime by coherent spin rotation, a larger value of $K_c^\parallel/M$ corresponds to diminished rotation of the magnetization vector per unit field. Returning to Figures 5 and 6, it appears that the experimental hysteresis loops would be better described by a value of $K_c^\parallel/M$ that is approximately a factor of two smaller. However, such a small value of the in-plane cubic anisotropy field is not in agreement with FMR results. We attribute this discrepancy to the small total sample magnetic moment ($\sim 10^{-7}$ emu) that is measured by the SQUID magnetometry when magnetization reversal is occurring by coherent spin rotation. This causes the data fitting algorithm to perform rather poorly, leading to significant error in the quoted value of the magnetic moment. On the other hand, the angular-dependence of the FMR is not subject to such limitations, which makes it better suited for the determination of the in-plane cubic anisotropy field than explicitly matching the slope of the measured and simulated field-dependent SQUID measurements. In light of these considerations, the agreement of simulation and experiment is quite reasonable.

Each half of the double hysteresis loop is centered at a field of roughly $K_u^{011}/M$. In actuality the switching angles of the noncoherent spin flips are influenced by the cubic anisotropy field, which leads to some deviation from this rule. As an example, for $y=0.010$ the simulated loops are centered about $|\mu_0 H|=4.1$ mT while $K_u^{011}/M \sim 3.5$ mT, a difference of about 20%. This effect becomes more pronounced as the magnitude of $K_c^\parallel/M$ decreases as the increased slope of the linear portion of $M$ vs. $H$ enhances this angular disparity resulting in larger shifts of the loop center. In the current work, this effect will be minor as changes in the magnitude of $K_c^{\parallel}/M$ are relatively small.

With these basic trends in mind the emergence of double hysteresis loops from kinked hysteresis loops when $H\parallel[011]$ can be easily understood in the context of the present model. For $y=0.010$ the hysteretic width of each "lobe" of the double loop, $\mu_0H_2-\mu_0H_1$, is ~2 mT. Each double loop is centered at approximately 4.1 mT. Therefore, $\mu_0H_1$, the field at which the first noncoherent flip occurs, is ~3 mT. The two halves of the double hysteresis loop do not "overlap" one another and are connected by a reversible linear region, resulting in the wasp-waisted lineshape. When $y=0$, however, $\mu_0H_2-\mu_0H_1$, is ~4 mT, owing to the increased value of $\Delta E^{011}$, while the center of each "lobe" has decreased to ~2.5 mT due mainly to the decrease in $K_u^{011}$. Hence, $\mu_0H_1\approx0$ resulting in an effective overlap of the two lobes, which leads to the "kinked" lineshape. Similarly, when $y$ is further increased to 0.021 and 0.027 from 0.010, the combined effect of decreasing $\Delta E^{011}$ and increasing $K_u^{011}/M$ leads to further wasp-waisting of the loops; the proportion of magnetization reversal that occurs by coherent spin rotation increases with respect to noncoherent spin switching. Consequently, the progression of [011]-oriented hysteresis loops as a function of $y$ is explained by the combined decrease of $\Delta E$ and increase in the field at which the double hystersis loop lobes are centered due to the increase of $K_u^{011}/M$.

## IV. CONCLUSIONS

In summary, we have investigated the effect of dilute alloying of the anion sublattice with S on in plane magnetic anisotropy and magnetization reversal in $Ga_{1-x}Mn_xP$. The in-plane unaxial anisotropy field along [011], $K_u^{011}$, can be controlled by adjusting the S concentration whereby the $\left[0\bar{1}1\right]$ axis is increasingly preferred as *y* increases. In parallel, the magnitude of the (negative) in-plane cubic anisotropy field, $K_{c1}^{\parallel}$, decreases as *y* increases. The interplay of these two trends creates an intricate free energy landscape in which magnetization reversal occurs by a combination of coherent spin rotation and noncoherent spin switching, which produces either a kinked or double hysteresis loop when the applied field is parallel to [011]. The occurrence of double hysteresis loops for larger values of *y* is facilitated by a decrease in the barrier to domain nucleation and propagation, $\Delta E^{011}$, which corresponds physically to a decrease in hysteretic losses as *y* is increased. Indeed, for *y*=0.027 the field-dependence of the magnetization is nearly completely reversible, in contrast to the behavior observed at smaller values of *y*. The unique hysteretic behavior presented herein is, therefore, the product of a multifaceted interplay of anisotropy and thermodynamic parameters, each of which has its own unique compositional dependence.

We attribute the changes in the magnetization reversal process to a decrease in *p* due to compensation by S donors, though note that we cannot directly exclude other possibilities. One possibility that we *can* exclude is that the observed changes in the magnetic anisotropy are due to the additional compressive strain associated with S substituting for P. We examined this possibility from results obtained from a parallel series of $Ga_{0.959}Mn_{0.041}P_{1-y}As_y$ films. When isovalent As substitutes for P, it adds compressive strain without intentional compensation. As was the case for $Ga_{0.959}Mn_{0.041}P_{1-y}S_y$, all $Ga_{0.959}Mn_{0.041}P_{1-y}As_y$ samples have $\left[0\bar{1}1\right]$ easy axes in the plane of the film. However, the anisotropy fields as determined from FMR were not altered by the presence of As on the Group V sublattice. Furthermore, all [011]-oriented hysteresis loops retained the kinked linkshape; no double hysteresis loops were observed for *y*<0.03. Since direct, accurate measurement of $p$ is not straightforward, somewhat indirect evidence in combination with the previously-discussed development of a paramagnetic resonance signal, decrease in $T_C$ and XMCD asymmetry with increasing $y$, is crucial to establishing that the results presented herein are truly due to the modulation of $p$ by compensating sulphur donors.

As a final point we comment on the remarkable similarity between the trends in the in-plane uniaxial anisotropy with $p$ for $Ga_{1-x}Mn_xAs$ and $Ga_{1-x}Mn_xP$. In fact, all magnetic anisotropies probed to date in $Ga_{1-x}Mn_xP$ have exhibited similar, if not identical, behavior to those observed in $Ga_{1-x}Mn_xAs$ [11, 12]. Further exploration of the magnetic anisotropy in $Ga_{1-x}Mn_xP$ as a function of $x$, $p$ and temperature is necessary to provide additional verification of these trends. Regardless, our collective work so far indicates that theories pertaining to the origin of the various magnetic anisotropies in $III_{1-x}Mn_xV$ must account for the fact that localized, impurity band holes are capable of mediating the same anisotropic interactions as the holes in metallic $Ga_{1-x}Mn_xAs$ [10].

**ACKNOWLEDGEMENTS**

Materials synthesis and SQUID magnetometry experiments at Lawrence Berkeley National Laboratory were supported by the Director, Office of Science, Office of Basic Energy Sciences, Division of Materials Sciences and Engineering, of the U.S. Department of Energy under Contract No. DE-AC02-05CH11231. The FMR work at the Walter Schottky Institut was supported by Deutsche Forschungsgemeinschaft through SFB 631 and the Bavaria California Technology Center. P.R.S. acknowledges support from NSF and NDSEG fellowships.

**Table 1** – Selected compositional parameters of $Ga_{1-x}Mn_xP_{1-y}S_y$ as determined by SIMS and ion beam analysis.

| Mn$^+$ Implant Dose (cm$^{-2}$) | $x$ | S$^+$ Implant Dose (cm$^{-2}$) | $y$ |
|---|---|---|---|
| 1.5x10$^{16}$ | 0.042 | 0 | 0 |
| 1.5x10$^{16}$ | 0.041 | 2.5x10$^{15}$ | 0.010±0.001 |
| 1.5x10$^{16}$ | 0.041 | 5.0x10$^{15}$ | 0.021±0.0015 |
| 1.5x10$^{16}$ | 0.041 | 7.3x10$^{15}$ | 0.027±0.002 |

TABLE 1 – Stone *et al*.

**Table 2** – Cubic and uniaxial anisotropy fields of $Ga_{1-x}Mn_xP_{1-y}S_y$ as determined by FMR.

| $y$ | $2K_u^{011}/M$ (mT) | $2K_c^{\parallel}/M$ (mT) | $2K_{eff}^{100}/M$ (mT) | $2K_c^{\perp}/M$ (mT) |
|---|---|---|---|---|
| 0 | 5±1 | -35±2 | 175±3 | -80±3 |
| 0.010 | 8±1 | -28±2 | 68±3 | -40±3 |
| 0.021 | 10.5±1.5 | -28±2 | 98±3 | -50±3 |
| 0.027 | 12.5±2 | -25±2 | 75±3 | -40±3 |

TABLE 2 – Stone *et al.*

**Table 3** – $\Delta E^{011}$ as a function of y. $\Delta E^{011}_{min}$ ($\Delta E^{011}_{max}$) are the minimum (maximum) value of $\Delta E^{011}$ for which the simulations reasonably describe the $M(H)$ loop.

| y | $\Delta E^{011}_{min}$ (x$10^{-4}$ meV/Mn) | $\Delta E^{011}_{max}$ (x$10^{-4}$ meV/Mn) |
|---|---|---|
| 0 | 1.5 | 3.1 |
| 0.010 | 0.95 | 1.9 |
| 0.021 | 0.62 | 1.2 |
| 0.027 | 0.19 | 0.62 |

TABLE 3 – Stone *et al.*

FIGURE CAPTIONS

**Figure 1** – Field dependence of the ferromagnetic resonance for $Ga_{0.958}Mn_{0.042}P_{1-y}S_y$ with $y=0$, $y=0.010$, $y=0.021$, and $y=0.027$ taken with the field applied parallel to the in-plane $[0\bar{1}\bar{1}]$ direction at $T=5K$. The magnetic field corresponding to paramagnetic $Mn^{2+}$ moments with $g=2.013$ is indicated by the dashed vertical line.

**Figure 2** – Mn (solid lines) and S (dashed lines) concentrations as a function of depth for samples with $y=0.010$ (black lines) and $y=0.027$ (grey lines) as determined by secondary ion mass spectrometry.

**Figure 3** – Dependence of the resonance field on the orientation of the applied magnetic field in the plane of the sample at $T=5K$ for (a) $y=0.010$ (b) $y=0.021$ and (c) $y=0.027$. Filled circles correspond to experimental data points while the solid lines are the results of FMR simulations. The missing data points in (c) are due to an overlap of the ferromagnetic and the paramagnetic resonance, which hampered evaluation of the resonance field. (d) The coordinate system used for FMR and SQUID simulations. The [100] direction is normal to the thin film plane.

**Figure 4** – Field dependence of the magnetization for (a) $y=0$, (b) $y=0.010$, (c) $y=0.021$, and (d) $y=0.027$ for $H\|[0\bar{1}\bar{1}]$ (dashed lines) and $H\|[011]$ (solid lines). Measurements were performed at $T=5K$ after saturating the magnetic moment at $\mu_0H=5T$. For further information regarding the calculation of the magnetic moment per $Mn_{Ga}$ refer to Appendix B of Ref. [12].

**Figure 5** – (a) Comparison of simulated and experimental hysteresis loops for $y=0.010$ for a single value of $\Delta E^{011}$. (b) Free energy as a function of the orientation of the magnetic moment in the plane of the sample ($\Theta$) at selected magnetic field strengths. Filled symbols correspond to the orientation of the magnetic moment at a given field. Open symbols and arrows represent noncoherent spin switching from one magnetization orientation to another which occurs at the specified field.

**Figure 6** – (a) Comparison of simulated and experimental hysteresis loops for $y=0.027$ for a single value of $\Delta E^{011}$. (b) Free energy as a function of the orientation of the magnetic moment in the plane of the sample ($\Theta$) at selected magnetic field strengths.

**Figure 7** – The effect of the in-plane anisotropy fields, $K_u^{011}/M$ and $K_c^{\parallel}/M$, and $\Delta E^{011}$ on the shape of the calculated double hysteresis loops. Black arrowheads along the solid $M(H)$ curve represent the progression of magnetization reversal. The orientation of the magnetic moment at magnetic fields specified by filled black circles is indicated by black arrows above or below the symbols with respect to the coordinate system included. The magnetic fields $\mu_0 H_1$ and $\mu_0 H_2$ correspond to those at which the first and second noncoherent spin flips occur. A positive magnetic field is defined as parallel to [011].

**Figure 8** – (main panel) Comparison of simulated and experimental hysteresis loops for $y=0.010$ in which $\Delta E^{011}$ is assumed to follow a Gaussian distribution with a mean of $1.2 \times 10^{-4}$ meV/Mn

and standard deviation of 4.9x10$^{-5}$ meV/Mn. (inset) Relative abundance of different $\Delta E^{011}$ according to the Gaussian distribution.

**Figure 9** – $\Delta E$ as a function of *y* for both [011] and $[0\bar{1}1]$ magnetization reversal processes. Symbols represent the mean value and the errors bars one standard deviation of $\Delta E$ within a Gaussian distribution.

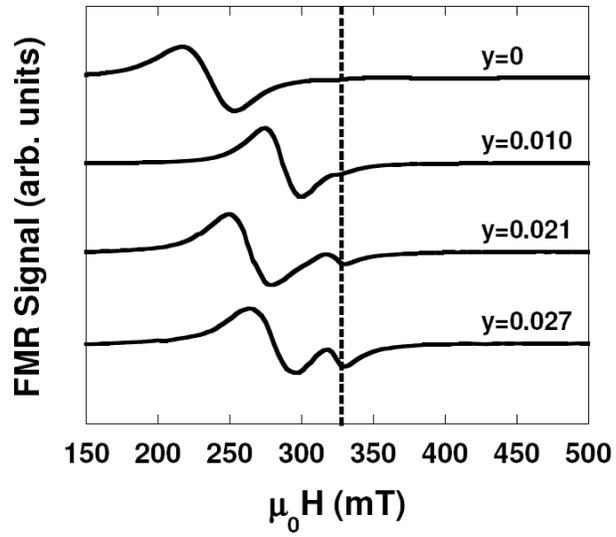

FIGURE 1- Stone *et al.*

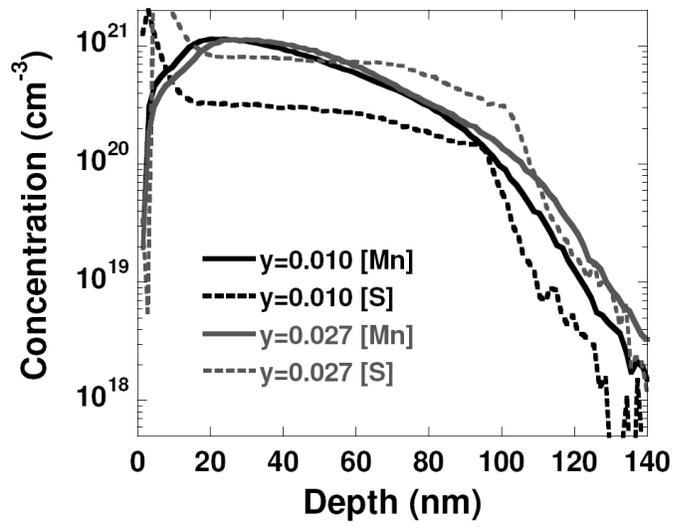

FIGURE 2- Stone *et al.*

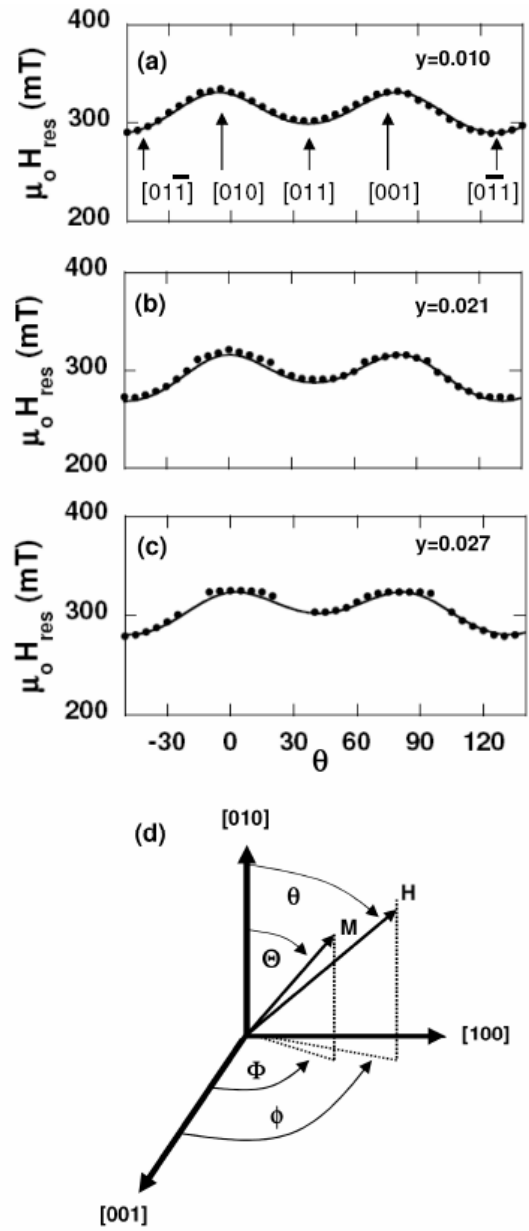

FIGURE 3- Stone *et al*.

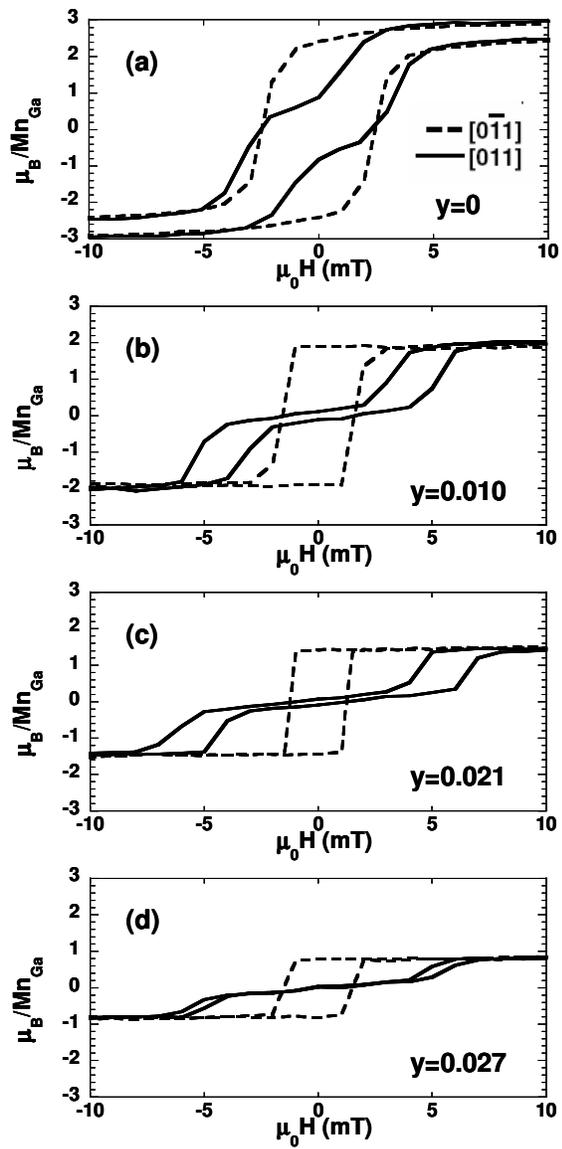

FIGURE 4 – Stone *et al.*

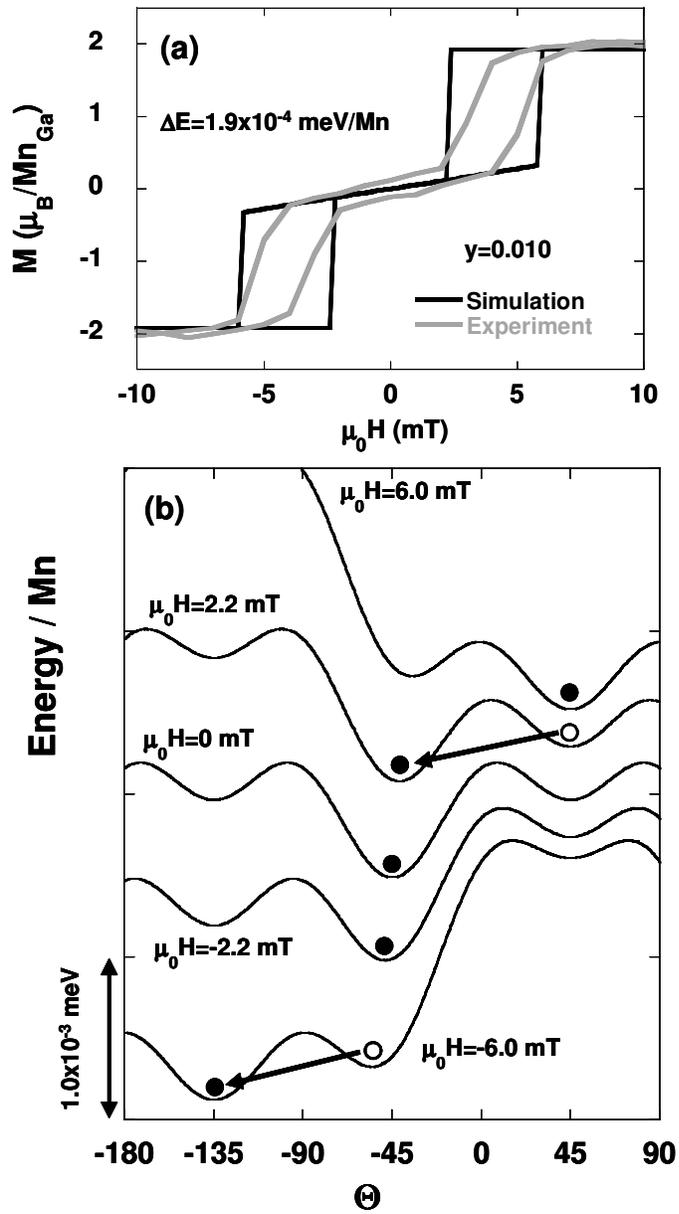

FIGURE 5 – Stone *et al.*

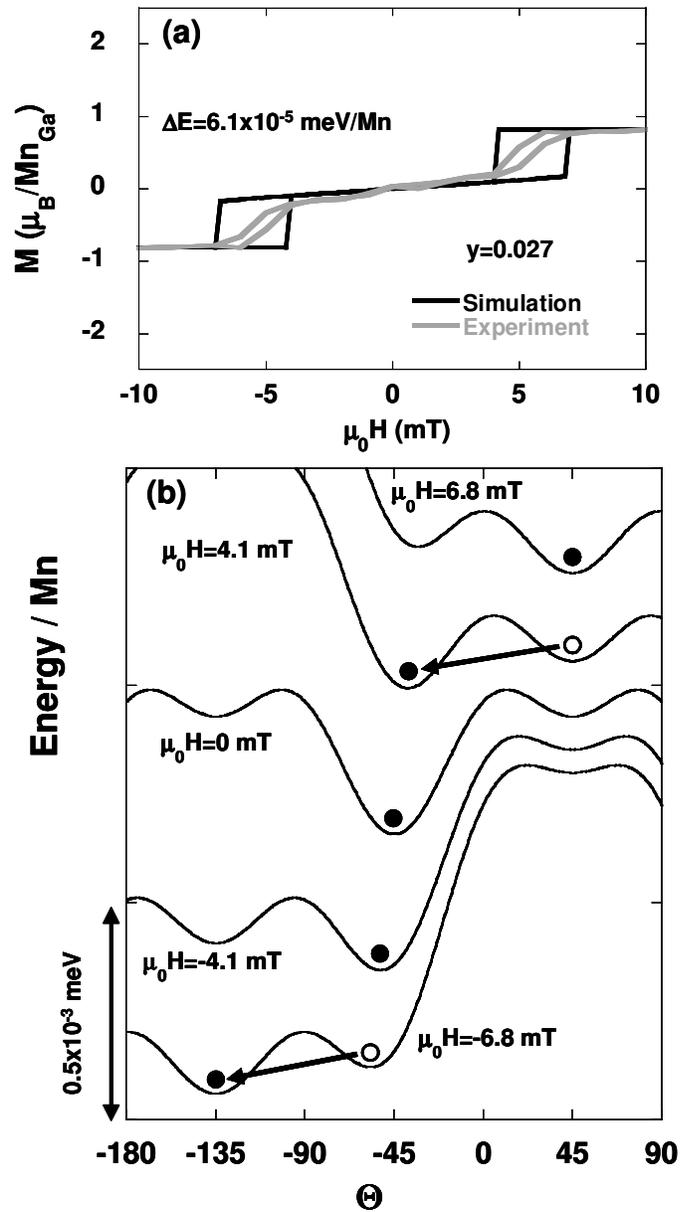

FIGURE 6 – Stone *et al.*

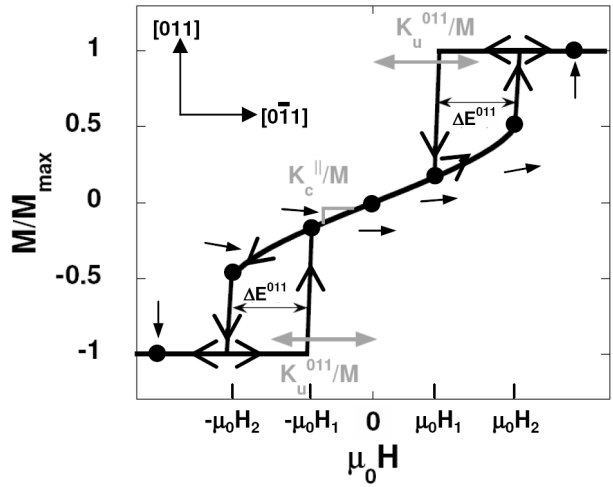

FIGURE 7 – Stone *et al*.

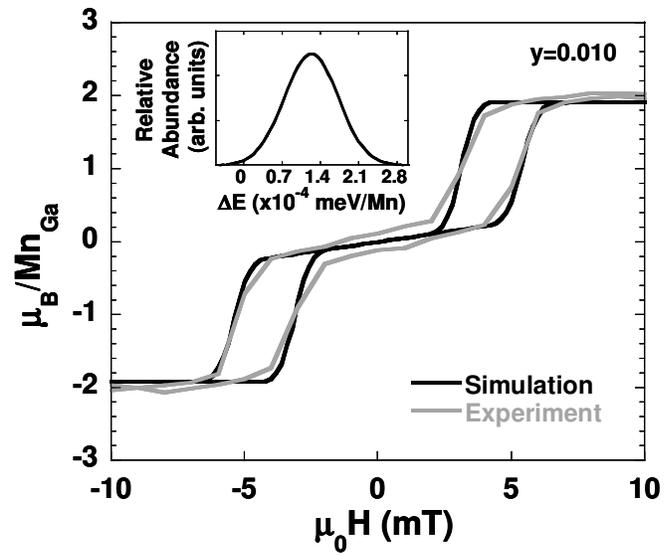

Figure 8 – Stone *et al.*

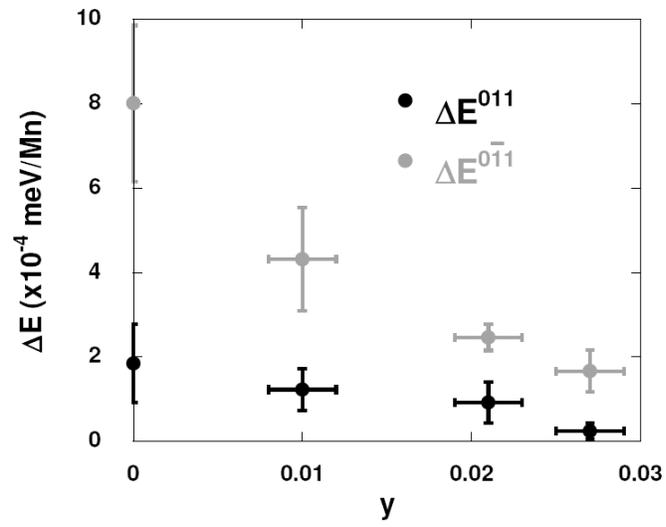

Figure 9 – Stone *et al.*